\documentclass[12pt, preprint]{aastex}

\begin{document}

\newcommand{\Eiso}{E_{\rm iso}}
\newcommand{\Egamma}{E_{\gamma}}
\newcommand{\Egiso}{E_{\gamma,{\rm iso}}}
\newcommand{\ERiso}{E_{R,{\rm iso}}}
\newcommand{\FluR}{S_R}
\newcommand{\FluGamma}{S_\gamma}
\newcommand{\PLidx}{\alpha}
\newcommand{\PHidx}{\Gamma_{\rm ph}}
\newcommand{\NH}{N_H}
\newcommand{\cosmology}{$H_0=65$ km s$^{-1}$ Mpc$^{-1}$, $\Omega_\Lambda = 0.7$ and $\Omega_m = 0.3$}

\shorttitle{The naked-eye burst GRB 080319B}
\shortauthors{Wo\'zniak et al.}

\title{Gamma-Ray Burst at the extreme: ``the naked-eye burst''\\
GRB 080319B}

\author{
P. R. Wo\'zniak,
W. T. Vestrand,
A. D. Panaitescu,
J. A. Wren,
H. R. Davis,
\& R. R. White
}

\affil{Los Alamos National Laboratory, MS-D466, Los Alamos, NM 87545 \\
email: (wozniak, vestrand, alin, jwren, hdavis, rwhite)@lanl.gov}

\begin{abstract}
On 19 March 2008, the northern sky was the stage of a spectacular optical transient that for a few seconds
remained visible to the naked eye. The transient was associated with GRB 080319B, a gamma-ray burst
at a luminosity distance of about 6 Gpc (standard cosmology), making it the most luminous optical object
ever recorded by human kind.
We present comprehensive sky monitoring and multi-color optical follow-up observations of GRB 080319B
collected by the RAPTOR telescope network covering the development of the explosion and the afterglow
before, during, and after the burst. The extremely bright prompt optical emission revealed features
that are normally not detectable. The optical and gamma-ray variability during the explosion are correlated,
but the optical flux is much greater than can be reconciled with single emission  mechanism and a flat
gamma-ray spectrum. This extreme optical behavior is best understood as synchrotron self-Compton model (SSC).
After a gradual onset of the gamma-ray emission, there is an abrupt rise 
of the prompt optical flux suggesting that variable self-absorption dominates the early optical light curve. 
Our simultaneous multi-color optical light curves following the flash show spectral evolution consistent with 
a rapidly decaying red component due to large angle emission and the emergence of a blue forward shock component 
from interaction with the surrounding environment. While providing little support for the reverse shock that 
dominates the early afterglow, these observations strengthen the case for the universal role of the SSC 
mechanism in generating gamma-ray bursts.

\end{abstract}

\keywords{gamma rays: bursts -- cosmology: observations -- shock waves}

\section{Introduction}
\label{sec:intro}

The current theoretical picture of the gamma-ray burst (GRB) phenomenon
involves a collapse of a massive, rotating star at the end of its normal evolution,
leading to the formation of a black hole (e.g. \citealt{woo06}). In the standard expanding fireball model
of GRBs part of the collapsar energy is channeled into a narrow ultra-relativistic
jet with a few degree opening angle and accelerated to Lorentz factor $\sim$100--1000 (e.g. \citealt{zha03}).
The progress in understanding the detailed geometry and energetics of the explosion has been greatly
stimulated by the launch of the Swift satellite (\citealt{geh04}) in the fall of 2004.
High precision rapid localizations from the BAT instrument (c.f. \citealt{sak08})
combined with new developments in fast optical follow-up and wide field monitoring technology
(\citealt{ake03,boe04,cov04,gui06,per04,rei05,ves02,bes08}) have enabled routine observations of the early
optical afterglow emission thought to arise in the external shock from the interaction of the jet
with the circum-burst medium. In a few bursts it was possible to detect the prompt optical emission
from internal shocks within the jet. (e.g. \citealt{blu06,ves05,ves06,ryk05}).

The intrinsic luminosity of GRBs spans several orders of magnitude, and studying
the extreme cases provides a handle on the source of this diversity.
The optical transient associated with GRB 080319B set a new record of both the apparent magnitude
and the intrinsic luminosity. For a few seconds it was detectable with the unaided eye from a dark sky site.
More importantly, new generation optical sky sentinels and rapid response telescopes
were able to measure---over an unprecedented range of brightness---the development of the explosion
and the afterglow in the minutes before, during, and after the stellar collapse.
Sampling the broad-band spectral evolution of GRBs and their afterglows with high-cadence, multi-wavelength
light curves reaching into the critical first minutes of the explosion is crucial to understanding
the emission mechanism. Here we present comprehensive sky monitoring  and multi-color optical follow-up observations
of GRB 080319B collected by the RAPTOR telescope network. 

\section{Observations}
\label{sec:data}

On March 19, 2008, at 06:12:49 Universal Time (UT) the Burst Alert Telescope (BAT) onboard
the Swift satellite (\citealt{geh04}) was triggered by an intense pulse of gamma rays from GRB 080319B (\citealt{rac08a}).
The BAT localization distributed over the Gamma-Ray Burst Coordinate Network (GCN) at 06:12:56 UT was
received by the RAPTOR telescope network within a second, while the follow-up response for
the previous alert (GRB 080319A) was still in progress. 
The system owned by the Los Alamos National Laboratory
is located at the Fenton Hill Observatory at an altitude of 2,500 m (Jemez Mountains, New Mexico). RAPTOR-Q
is a continuous all sky monitor with the approximate magnitude limit 9.5 (unfiltered $R$-band equivalent, $C_{\rm R}$).
RAPTOR-P is an array of four 200-mm Canon telephoto lenses, each covering a field of view $8\times8$ square degrees
down to $C_{\rm R} \sim 15$ mag, normally used to patrol large areas of the sky for optical flashes. RAPTOR-T instrument
consists of four co-aligned 0.4-m telescopes on a single fast-slewing mount and provides simultaneous images
in four photometric bands ($V$, $R$, $I$ and clear). Both rapidly slewing telescopes available
on the network, RAPTOR-P and T, responded in the override mode upon receiving the X-ray Telescope
(XRT) localization at 06:13:16 UT with the first follow-up exposure starting at 06:13:24 UT
(35 s after the BAT trigger). Independently of the follow-up observations, the burst location
was covered by our continuous all-sky monitor RAPTOR-Q. Those measurements rule out the presence
of optical precursor for 30 minutes prior to the gamma-ray burst down to a flux limit of 0.5 Jy ($3\sigma$)
or just 2.5\% of the peak flux. Unfiltered RAPTOR-Q and P magnitudes were calibrated
using several hundred stars from the Northern Sky Variability Survey (NSVS; \cite{woz04}).
The color terms and absolute calibration for RAPTOR-T measurements were obtained with
1--2 dozen stars per filter using secondary field standards from \cite{hen08}.
Observations were collected in sub-optimal weather with wind-shake affecting about 50\% of RAPTOR-T measurements.
The early non-detections in RAPTOR-T data are due to saturation.
Photometric measurements after $t = 2200$ s are based on co-added images utilizing between 5 and 20 individual exposures.
The combined error bars are photon noise estimates with the systematic uncertainty of 0.02 mag added in quadrature.

\section{Results}
\label{sec:results}

\begin{figure}
\plotone{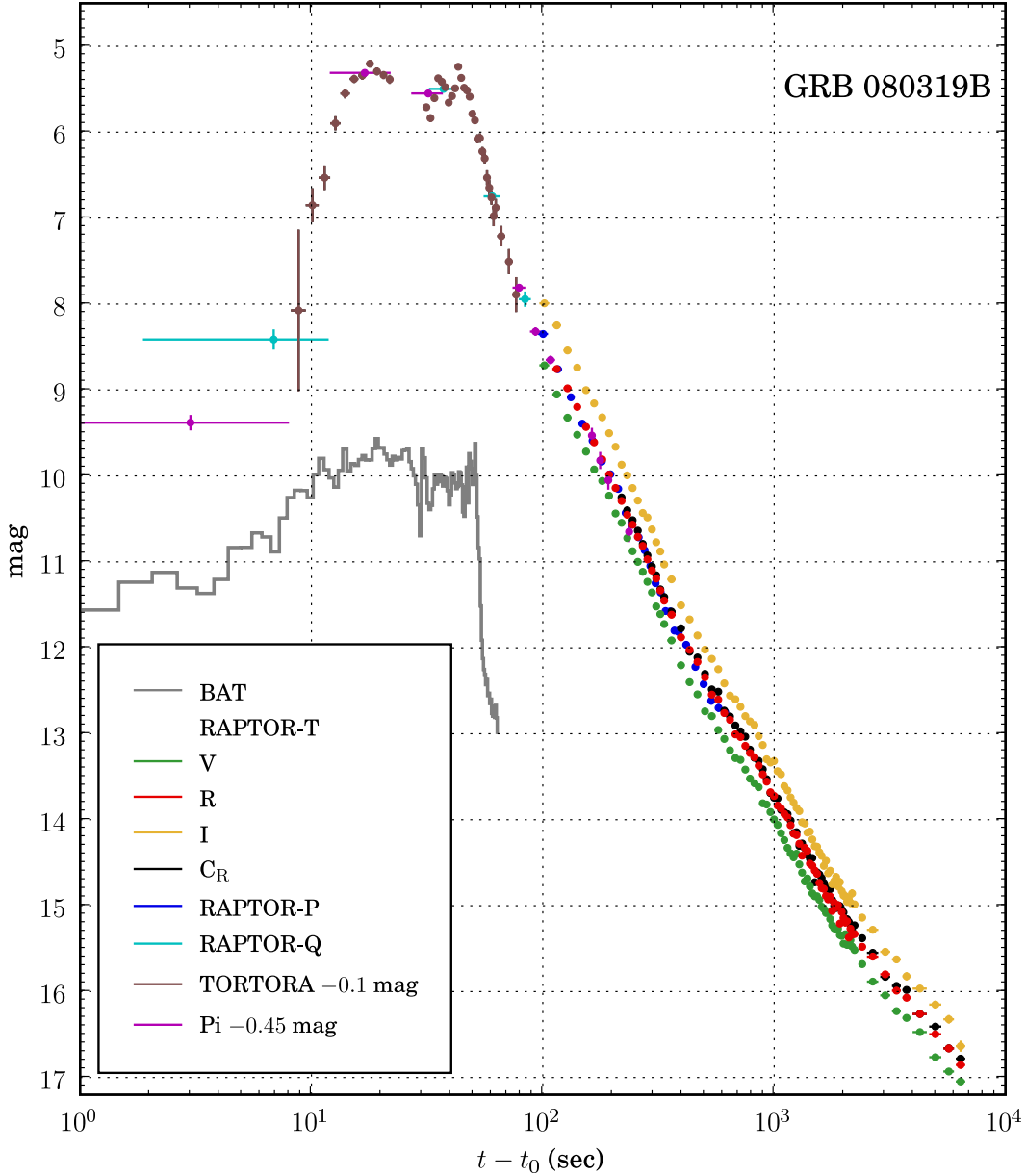}
\vspace{-2.0cm}
\figcaption[]{\label{fig:comb_lc}
Optical light curve of GRB 080319B, ``the naked eye burst''.
The transient was independently detected and followed
over a high dynamic range by three fully autonomous  instruments on the RAPTOR telescope network:
a simultaneous multi-color imager RAPTOR-T, a wide-field survey array RAPTOR-P, and an all-sky monitor
RAPTOR-Q. A non-detection in RAPTOR-Q data rules out the presence of a significant optical precursor for $\sim 30$ minutes
prior to BAT trigger at $t_0$. The RAPTOR measurements of the prompt optical emission during the gamma-ray burst
($t-t_0$ = 0--60 s.) agree with the data collected by other instruments reporting detections:
TORTORA (\citealt{kar08}) and Pi-of-the-sky (\citealt{cwi08}). Unfiltered observations were adjusted to $R$-band
equivalent scale.
}
\end{figure}

The light curve resulting from our coordinated measurements (Fig.~\ref{fig:comb_lc}, Table~\ref{tab:data})
shows the optical flash that within several seconds of the onset reached a peak visual magnitude
$\sim 5.3$. At red-shift $z = 0.937$ (\citealt{vre08}) and assuming standard cosmology
($\Omega_{\rm m} = 0.27, \Omega_\Lambda = 0.73, H_0 = 71$ km/s)
this corresponds to absolute magnitude $M_{\rm V} \simeq -38.6$ breaking the previous
luminosity record held by GRB 990123 by nearly an order of magnitude.
The simultaneous multi-color measurements recorded by RAPTOR-T starting at $t \simeq 100$ s (Figs.~\ref{fig:colors} and
\ref{fig:models}) reveal an afterglow light curve with a very steep initial decline ($\alpha_1 \sim 2.5$), modulated
by several small bumps, and followed by a gradual transition to a slower decay after 800--1000 s with $\alpha_2 \sim 1.2$.
The transition is accompanied by a change in color $\Delta(V-I) \simeq -0.25$ mag that is largely a result
of mixing between two components with colors that are approximately constant.
These trends agree with other published data on GRB 080319B from PAIRITEL, KAIT, Nickel, UVOT, and Gemini-S instruments
(\citealt{blo08}). \cite{rac08c} presented an extensive broad-band study of GRB 080319B based on several independent
data sets. All available data sets are consistent within the quoted uncertainties.
In our joined model fits (Fig.~\ref{fig:models} and Table~\ref{tab:models}) we included the data from \cite{blo08}
in order to constrain the host galaxy reddening and late time decay rate. 
We adopted a double power-law model with SMC-like extinction (\citealt{pei92}) and spectral index $\beta_1, \beta_2$
for the red and blue components with decay rates $\alpha_1, \alpha_2$ that dominate,
correspondingly, early and late times. There is only a modest amount of dust in the host galaxy reference frame,
$A_{\rm V} \simeq 0.06$ mag, and the Galactic extinction $E_{\rm B-V} = 0.01$ is negligible (\citealt{sch98}).
While the model with constant intrinsic spectral slopes
provides an acceptable fit to the data, the afterglow becomes redder again after $\sim 10^4$ s.
Adding a third power-law results in a degenerate fit, but we can approximately assess the effect of the additional component
by adding a time dependent spectral slope $\beta_2$ with linear dependence on $\log t$ that significantly improves the fit
(Table~\ref{tab:models}).

\section{Discussion}
\label{sec:discussion}

\subsection{Prompt optical emission}
\label{sec:prompt}

The variations of the optical emission during the burst roughly follow the onset, the three major peaks,
and the final decline of the gamma-ray flux (Fig. 2), indicating that both types of radiation have a closely
linked source. In the standard fireball model (\citealt{mes93,wax97}), this prompt emission is generated in internal
shocks due to colliding shells of material moving at different bulk Lorentz factors within the ultra-relativistic
outflow. From the burst spectrum measured by Konus satellite (\citealt{gol08}) we find the gamma-ray flux density
$\sim7$ mJy near the peak at 650 keV. The flux density observed in the optical band in the same time interval
is $\sim10$ Jy and falls four orders of magnitude above the extrapolation of the low-energy gamma-ray spectrum
($F \propto \nu^{0.18\pm0.01}$). The most natural scenario that accommodates the presence of these two distinct
spectral components is a synchrotron self-Compton (SSC) model, in which
optical photons are generated as synchrotron emission and gamma rays are a result of inverse-Compton
scatterings of those photons to higher energies. Both processes operate simultaneously
in the power-law distribution of electrons accelerated by the shock front, producing a tracking behavior
between the low-energy synchrotron hump and its high-energy ``Compton image''. The apparent degree of correlation
between the gamma-ray and optical light curves can vary widely due to several factors such as the presence
of the early afterglow emission, the effective width of the energy distribution of scattering electrons,
and the location of the compared frequency ranges within the double peaked SSC spectrum. Above the peak frequency,
a typical synchrotron spectrum decreases sharply as $\sim \nu^{\rm -p/2}$
with $p \sim 3$, sometimes faster, if the corresponding electrons cannot cool effectively (\citealt{pan07}).
Therefore, the observed prompt optical flux and its ratio to gamma-ray flux will be very sensitive to the location
of the synchrotron peak, especially if most of the low frequency $\nu^{1/3}$ shoulder is self-absorbed.
The extraordinary brightness of the prompt optical emission in GRB 080319B can then be understood as a result
of fortuitous positioning of the synchrotron peak in the optical band.

The rise of the prompt optical emission to its peak flux at $t \sim 18$ s is somewhat delayed
and extremely steep by comparison with the onset of the gamma-ray emission. We find
$F \propto t^\alpha$ with $\alpha_{\rm opt} = 3.08\pm0.20$
for TORTORA light curve and $\alpha_\gamma \simeq 1/3$ observed by BAT, so even after scaling
by the gamma-ray emission the optical flux rises as $\sim t^3$ prior to the first peak (Fig. 2).
This suggests that at the beginning of the burst $\nu_{\rm opt} < \nu_{\rm a}$,
i.e. the optical frequency is below the self-absorption cut-off of the synchrotron radiating fluid.
Assuming that the emitting shell of shocked material in the relativistic outflow expands and cools adiabatically,
we can estimate the evolution of the electron Lorentz factor $\gamma_{\rm e} \propto t^{-1}$
and magnetic field $B \propto t^{-2}$ with constant number of electrons $N_{\rm e}$ and bulk Lorentz factor
$\Gamma$. The synchrotron emissivity spectrum resulting from the power-law distribution of electrons
$N_{\rm e}(\gamma_{\rm e}) \propto \gamma_{\rm e}^{\rm -p}$ will then have a peak flux $F_\nu(\nu_{\rm p}) \propto t^{-2}$
at $\nu_{\rm p} \propto t^{-4}$, and the self-absorption frequency will decrease with time as $\nu_{\rm a} \propto t^{-1}$
for $\nu_{\rm p} < \nu_{\rm a}$, or $\nu_{\rm a} \propto t^{\rm -(4p+6)/(p+4)}$ otherwise. In this scenario,
the optical frequency must be between the peak of emissivity and the self-absorption cut-off
($\nu_{\rm p} < \nu_{\rm opt} < \nu_{\rm a}$) in order for the predicted optical light curve $F_\nu(\nu_{\rm opt}) \propto t^3$
to fit the observations. Other possible arrangements of frequencies result in a much slower flux increase
$F_\nu(\nu_{\rm opt}) \propto t$. The ``unveiling'' of the optical synchrotron peak would occur a few seconds after
the corresponding pulse of gamma rays, when the adiabatic expansion of the shocked medium had developed.
The model also predicts a very blue optical spectrum $F_\nu \propto \nu^{2.5}$ during the fast initial rise,
providing a well-defined observational test.

There is a growing evidence that the basic SSC mechanism is important during the prompt phase of most classical
long-duration GRBs. Optically bright bursts detected during the gamma-ray emission, such as GRB 080319B, 990123 and 061126,
tend to have a large ``excess'' of optical light compared to the continuation of the high energy component (\citealt{cor05,per08}),
consistent with the synchrotron peak falling near or slightly below 2 eV. On the other hand, in two well observed bursts
showing faint prompt optical emission (GRB 041219B and 050820A; $F_{\rm opt}/F_\gamma \sim 10^{-5}$)
the implied average optical-to-gamma spectrum is flatter than $\nu^{-1/2}$ and the optical and gamma-ray light curves are closely
correlated (\citealt{ves05,ves06}). Without additional information, such as the color of the prompt optical light, it is impossible
in this case to rule out a single spectral component peaking between 2 eV and 100 keV. However, in both objects the high-energy end
of the gamma-ray spectrum follows the optical flux more closely than the low-energy part, as expected in the SSC scenario
with the optical band well above the synchrotron peak.

\subsection{Transition to afterglow}
\label{sec:afterglow}

\begin{figure}
\plotone{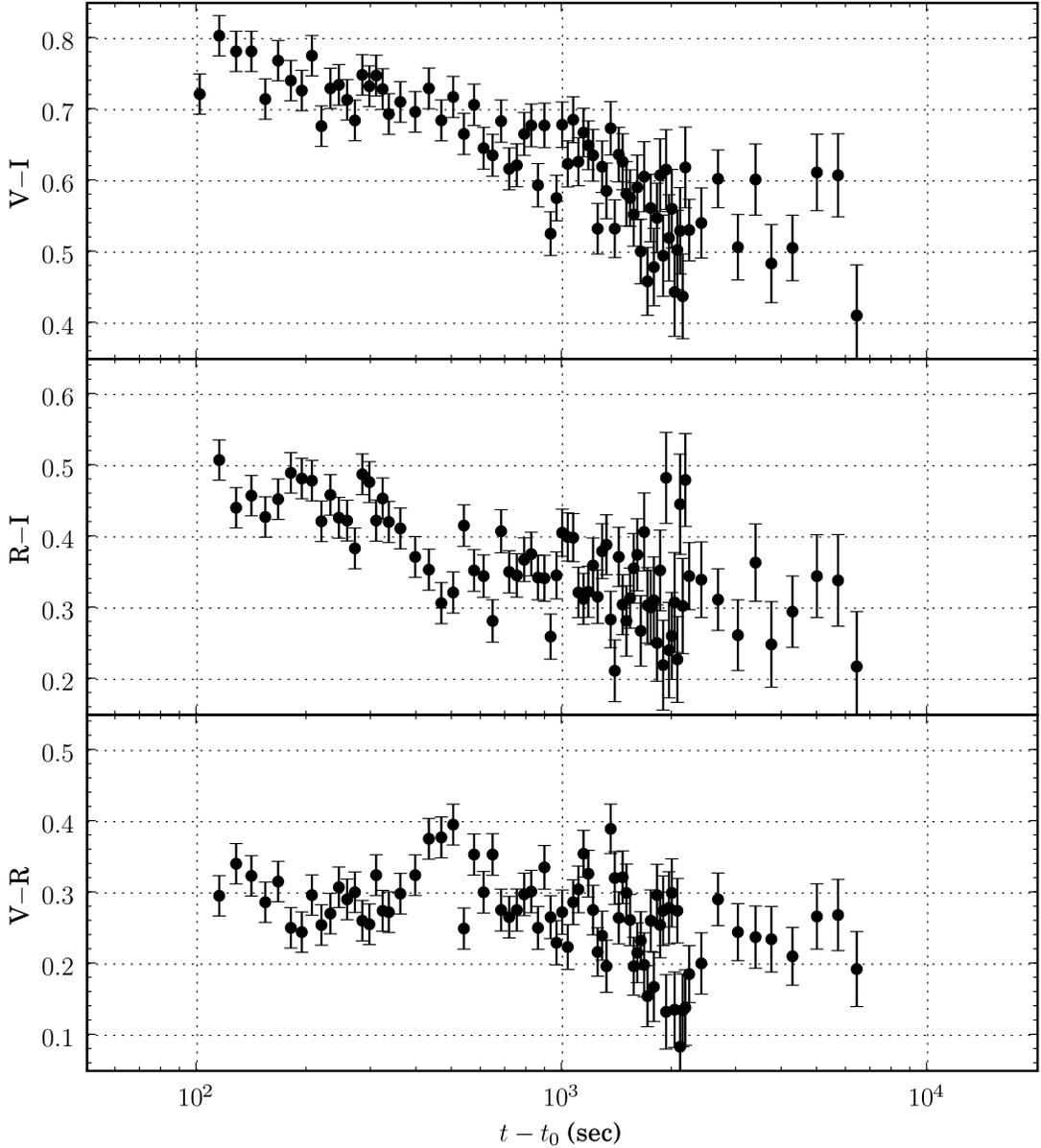}
\figcaption[]{\label{fig:colors}
Color evolution of the optical afterflow of GRB 080319B. The simultaneous multi-color measurements collected by RAPTOR-T
show that the apparent color of the optical transient gradually evolves from $(V-I) \simeq 0.80$ at 100 s
toward a much bluer value $(V-I) \simeq 0.55$ around 2000 s, and then remains approximately constant until 6500 s.
There is no indication of short time-scale variability in color despite the presence of several bumps in the light curve that modulate
smoothly decaying afterglow flux by about 10\% on time-scales 100--300 s.
}
\end{figure}

\begin{figure}
\vspace{-1.0cm}
\plotone{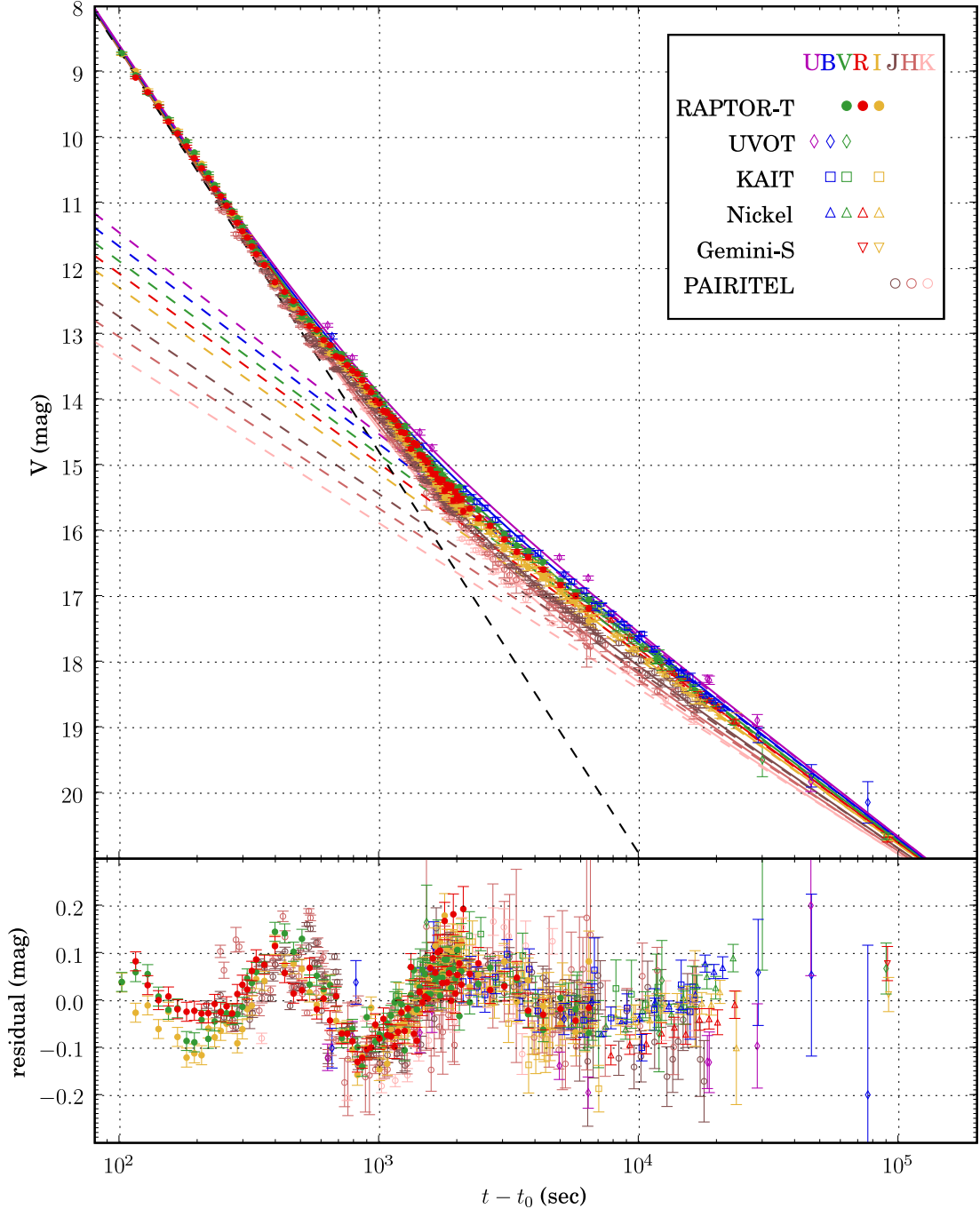}
\vspace*{-18mm}
\figcaption[]{\label{fig:models}
Optical and NIR light curves of the GRB 080319B afterglow. The simultaneous multi-color RAPTOR-T
measurements are compared to data from other instruments (\cite{blo08}). The lines show
the best fit model with color evolution from Table~\ref{tab:models}. Dashed lines are for individual
components and solid lines are the totals in each filter. The measurements and
the model fits in all photometric bands were shifted to absorb the constant color of the
early fast-decaying component using the $V$ band as reference.
}
\end{figure}

The parameters of the early fast-decaying component
are marginally consistent with the relation $\alpha = 2+\beta$ expected for large-angle emission that sets the upper limit on the rate
of flux decline due to photons emitted at angles larger than $1/\Gamma$ relative to the direction
toward the observer (\citealt{kum00}). The measurements are harder to reconcile with the reverse shock model (\citealt{kob00,pan04}) that predicts
a slower decay with $1.5 < \alpha < 2.0$ for the observed spectrum. This suggests that the early optical afterglow flux arises
from the prompt radiation mechanism and that those optical photons arrive at observer later than the burst simply due to
a longer photon path from the curved emitting surface to the observer. The contrast generated by exceptionally bright prompt
optical signal in GRB 080319B makes it easier to detect large-angle emission from the burst tail that competes with the slower decaying
external shock emission. Within the context of the standard fireball model, the transition to a shallower decay after 800 s
is most likely associated with the emergence of the forward shock driven by the relativistic ejecta into the circum-burst medium
(\citealt{mes97,sar98}).
However, assuming a constant color we obtain values $\alpha_2$ and $\beta_2$ that do not fit the standard blast wave solutions
for the most likely density profiles $r^{-2}$ (stellar wind) and $r^0$ (uniform interstellar medium). It is also difficult
to explain the X-ray light curve (\citealt{rac08b}) becoming steeper around 3000 s. At very late times ($\sim 4\times10^4$ s)
the model with evolving color approaches the closure relation $\alpha = (3\beta+1)/2$ for a wind medium (Table~\ref{tab:models}).
The data from other instruments point to similar difficulties with the canonical model (\citealt{blo08}, \citealt{rac08c}).
\cite{rac08c} show that a consistent interpretation of the XRT light curve and the blue color of the slow-decaying
component at early times can be achieved by extending the model with an additional strongly collimated outflow (two nested jets),
or introducing a complicated density profile in the external medium (local clumps). The specific model fit assuming the second scenario
presented by \cite{rac08c} suggests that the cooling break crossed the optical bands ($I \rightarrow R \rightarrow V$)
between $t = 150$ and 250 s. This would result in a dramatic change of the spectral slope $\Delta \beta \simeq -0.4$ ($\Delta (V-I) \simeq -0.25$ mag)
in just 100 s, contradicting the simultaneous color measurements from RAPTOR-T (Fig.~\ref{fig:colors}). The $(V-R)$ and $(R-I)$ color curves
in Fig.~\ref{fig:colors} do show some structure in excess of the systematic errors near $t \sim 500$ s, but the reality of this feature
is uncertain due to coinciding problems with wind-shake. On the other hand, the model fits in Table~\ref{tab:models}
provide a hint that a separate blue component with the temporal profile similar to the X-ray light curve contributes a modest
fraction of the total optical flux near $t \sim 1000$ s, before the slow decaying forward shock emission starts to dominate.
Therefore, the RAPTOR data are consistent with the preferred model of \cite{rac08c} comprising a narrow jet (half opening
angle $\theta_{\rm j} \simeq 0.2\arcdeg$) inside a less collimated outflow ($\theta_{\rm j} \simeq 4\arcdeg$),
and generating optical emission from forward shocks at two separate sites.

\subsection{Future outlook}
\label{sec:future}

The unprecedented panchromatic and temporal coverage of GRB 080319B revealed complexities in both the burst mechanism
and the resulting afterglow that are challenging the standard fireball model. Simultaneous multi-color observations
in optical/NIR energies during the burst proper are within the reach of the current rapid response telescopes for bright
and long-lasting events. With the introduction of even modest spectral capability---such as measuring the sign
of the spectral index---the next generation of persistent wide-field sky monitors (\citealt{ves02,bes08})
will provide a solid identification of the burst radiation mechanism and probe its possible diversity.

\acknowledgements

This research was performed as part of the Thinking Telescopes and RAPTOR projects supported by
the Laboratory Directed Research and Development (LDRD) program at LANL.

\begin{deluxetable}{rcccrcccrcccrcc}
\tablecaption{\label{tab:models}{Multi-color light curve models of GRB 080319B.}}
\tablehead{
\colhead{$\alpha_1$} &
\colhead{$\beta_1$} &
\colhead{$\alpha_2$} &
\colhead{$\beta_2$} &
\colhead{$A_{\rm V}$} &
\colhead{${\rm d}\beta_2 / {\rm d}\log t$} &
\colhead{$\chi^2_\nu(\chi^2/{\rm dof})$}
}

\startdata
$2.47^{+0.02}_{-0.02}$ &
$0.55^{+0.02}_{-0.04}$ &
$1.15^{+0.01}_{-0.01}$ &
$0.17^{+0.02}_{-0.06}$ &
$0.06^{+0.02}_{-0.06}$ &
\nodata                &
$1.04 (713.8/687)$ \\
$2.45^{+0.02}_{-0.02}$ &
$0.63^{+0.02}_{-0.04}$ &
$1.18^{+0.01}_{-0.01}$ &
$0.22^{+0.02}_{-0.12}$ &
$0.07^{+0.01}_{-0.07}$ &
$0.28^{+0.07}_{-0.07}$ &
$0.96 (656.3/686)$ \\
\enddata

\tablecomments{The light curve clearly shows small bumps and wiggles that are not captured by the model.
Measurement errors were rescaled to achieve $\chi^2_\nu \sim 1.0$ by adding a systematic
uncertainty 0.07 mag to all data points. The value of $\beta_2$ in the second model
with color evolution corresponds to $t = 10^4$ s.}

\end{deluxetable}

\begin{deluxetable}{rrcccc}
\tabletypesize{\small}
\tablecolumns{6}
\tablecaption{\label{tab:data}{RAPTOR observations of GRB 080319B.}}
\tablehead{
\colhead{\makebox[1.0cm][c]{$t_{\rm mid}$}} &
\colhead{\makebox[1.0cm][c]{$\Delta t_{\rm exp}$}} &
\colhead{\makebox[2.8cm][c]{$C_{\rm R}$}} &
\colhead{\makebox[2.8cm][c]{$V$}} &
\colhead{\makebox[2.8cm][c]{$R$}} &
\colhead{\makebox[3.0cm][c]{$I$}} \\
\colhead{\makebox[1.0cm][c]{(s)}} &
\colhead{\makebox[1.0cm][c]{(s)}} &
\colhead{(mag)} &
\colhead{(mag)} &
\colhead{(mag)} &
\colhead{(mag)}
}
\startdata
\cutinhead{RAPTOR-Q}
   6.87 &   10.00 &  $\phn8.412\pm0.118$ & \nodata & \nodata & \nodata \\
  37.22 &   10.00 &  $\phn5.497\pm0.017$ & \nodata & \nodata & \nodata \\
  60.48 &   10.00 &  $\phn6.747\pm0.035$ & \nodata & \nodata & \nodata \\
  83.81 &   10.00 &  $\phn7.943\pm0.089$ & \nodata & \nodata & \nodata \\
\cutinhead{RAPTOR-P}
 100.31 &   10.00 &  $\phn8.348\pm0.020$ & \nodata & \nodata & \nodata \\
 116.19 &   10.00 &  $\phn8.758\pm0.020$ & \nodata & \nodata & \nodata \\
 132.25 &   10.00 &  $\phn9.085\pm0.021$ & \nodata & \nodata & \nodata \\
 148.22 &   10.00 &  $\phn9.392\pm0.021$ & \nodata & \nodata & \nodata \\
 164.07 &   10.00 &  $\phn9.591\pm0.021$ & \nodata & \nodata & \nodata \\
 179.95 &   10.00 &  $\phn9.832\pm0.021$ & \nodata & \nodata & \nodata \\
 195.98 &   10.00 &  $\phn9.981\pm0.022$ & \nodata & \nodata & \nodata \\
 211.85 &   10.00 & $10.151\pm0.022$ & \nodata & \nodata & \nodata \\
 228.09 &   10.00 & $10.431\pm0.023$ & \nodata & \nodata & \nodata \\
 243.72 &   10.00 & $10.567\pm0.024$ & \nodata & \nodata & \nodata \\
 259.71 &   10.00 & $10.715\pm0.024$ & \nodata & \nodata & \nodata \\
 275.57 &   10.00 & $10.865\pm0.026$ & \nodata & \nodata & \nodata \\
 291.53 &   10.00 & $11.047\pm0.028$ & \nodata & \nodata & \nodata \\
 307.18 &   10.00 & $11.248\pm0.028$ & \nodata & \nodata & \nodata \\
 323.25 &   10.00 & $11.357\pm0.030$ & \nodata & \nodata & \nodata \\
 339.30 &   10.00 & $11.572\pm0.035$ & \nodata & \nodata & \nodata \\
 355.19 &   10.00 & $11.576\pm0.035$ & \nodata & \nodata & \nodata \\
 370.71 &   10.00 & $11.800\pm0.039$ & \nodata & \nodata & \nodata \\
 386.57 &   10.00 & $11.817\pm0.039$ & \nodata & \nodata & \nodata \\
 416.62 &   30.00 & $11.965\pm0.030$ & \nodata & \nodata & \nodata \\
 456.57 &   30.00 & $12.222\pm0.033$ & \nodata & \nodata & \nodata \\
 495.99 &   30.00 & $12.423\pm0.038$ & \nodata & \nodata & \nodata \\
 534.69 &   30.00 & $12.617\pm0.043$ & \nodata & \nodata & \nodata \\
 575.05 &   30.00 & $12.699\pm0.047$ & \nodata & \nodata & \nodata \\
 614.77 &   30.00 & $12.751\pm0.047$ & \nodata & \nodata & \nodata \\
\cutinhead{RAPTOR-T}
 101.63 &   10.00 & \nodata & $\phn8.713\pm0.020$ & \nodata             & $\phn7.991\pm0.020$ \\
 114.96 &   10.00 & \nodata & $\phn9.052\pm0.020$ & $\phn8.756\pm0.020$ & $\phn8.248\pm0.020$ \\
 127.89 &   10.00 & \nodata & $\phn9.322\pm0.020$ & $\phn8.981\pm0.020$ & $\phn8.540\pm0.020$ \\
 140.81 &   10.00 & \nodata & $\phn9.521\pm0.020$ & $\phn9.197\pm0.020$ & $\phn8.739\pm0.020$ \\
 153.74 &   10.00 & \nodata & $\phn9.717\pm0.020$ & $\phn9.430\pm0.020$ & $\phn9.002\pm0.020$ \\
 166.66 &   10.00 & \nodata & $\phn9.925\pm0.020$ & $\phn9.609\pm0.020$ & $\phn9.156\pm0.020$ \\
 180.60 &   10.00 & \nodata &    $10.059\pm0.020$ & $\phn9.808\pm0.020$ & $\phn9.318\pm0.020$ \\
 193.52 &   10.00 & \nodata &    $10.230\pm0.020$ & $\phn9.985\pm0.020$ & $\phn9.503\pm0.020$ \\
 206.25 &   10.00 & \nodata &    $10.437\pm0.020$ &    $10.140\pm0.020$ & $\phn9.661\pm0.020$ \\
 218.97 &   10.00 & $10.250\pm0.020$ & $10.546\pm0.020$ & $10.291\pm0.020$ & $\phn9.869\pm0.020$ \\
 231.70 &   10.00 & $10.399\pm0.020$ & $10.723\pm0.020$ & $10.452\pm0.020$ & $\phn9.993\pm0.020$ \\
 244.43 &   10.00 & $10.515\pm0.020$ & $10.876\pm0.020$ & $10.568\pm0.020$ & $10.141\pm0.020$ \\
 257.45 &   10.00 & $10.637\pm0.020$ & $11.000\pm0.020$ & $10.709\pm0.020$ & $10.286\pm0.020$ \\
 270.38 &   10.00 & $10.791\pm0.020$ & $11.116\pm0.020$ & $10.815\pm0.020$ & $10.431\pm0.020$ \\
 283.30 &   10.00 & $10.925\pm0.020$ & $11.232\pm0.020$ & $10.971\pm0.020$ & $10.483\pm0.020$ \\
 296.23 &   10.00 & $11.053\pm0.020$ & $11.356\pm0.020$ & $11.100\pm0.020$ & $10.623\pm0.020$ \\
 309.15 &   10.00 & $11.156\pm0.020$ & $11.519\pm0.020$ & $11.194\pm0.020$ & $10.771\pm0.020$ \\
 322.08 &   10.00 & $11.320\pm0.020$ & $11.608\pm0.020$ & $11.333\pm0.020$ & $10.879\pm0.020$ \\
 334.80 &   10.00 & $11.409\pm0.020$ & $11.724\pm0.020$ & $11.451\pm0.020$ & $11.030\pm0.020$ \\
 360.05 &   30.00 & $11.579\pm0.020$ & $11.914\pm0.020$ & $11.615\pm0.020$ & $11.203\pm0.020$ \\
 395.50 &   30.00 & $11.773\pm0.020$ & $12.203\pm0.020$ & $11.878\pm0.020$ & $11.506\pm0.020$ \\
 431.04 &   30.00 & $12.046\pm0.020$ & $12.400\pm0.020$ & $12.024\pm0.020$ & $11.670\pm0.020$ \\
 466.29 &   30.00 & $12.113\pm0.020$ & $12.542\pm0.020$ & $12.164\pm0.020$ & $11.857\pm0.020$ \\
 502.44 &   30.00 & $12.303\pm0.020$ & $12.739\pm0.020$ & $12.343\pm0.020$ & $12.021\pm0.021$ \\
 537.68 &   30.00 & $12.484\pm0.020$ & $12.795\pm0.020$ & $12.545\pm0.021$ & $12.129\pm0.021$ \\
 573.23 &   30.00 & $12.511\pm0.020$ & $12.956\pm0.020$ & $12.602\pm0.021$ & $12.249\pm0.021$ \\
 608.77 &   30.00 & $12.731\pm0.020$ & $13.060\pm0.020$ & $12.759\pm0.021$ & $12.414\pm0.021$ \\
 644.32 &   30.00 & $12.799\pm0.020$ & $13.192\pm0.020$ & $12.838\pm0.021$ & $12.556\pm0.021$ \\
 679.86 &   30.00 & $12.904\pm0.020$ & $13.283\pm0.020$ & $13.007\pm0.021$ & $12.599\pm0.022$ \\
 715.21 &   30.00 & $12.972\pm0.020$ & $13.304\pm0.020$ & $13.038\pm0.021$ & $12.687\pm0.021$ \\
 750.45 &   30.00 & $13.034\pm0.020$ & $13.417\pm0.021$ & $13.141\pm0.021$ & $12.795\pm0.022$ \\
 786.09 &   30.00 & $13.185\pm0.020$ & $13.522\pm0.021$ & $13.224\pm0.022$ & $12.856\pm0.022$ \\
 821.85 &   30.00 & $13.280\pm0.021$ & $13.576\pm0.021$ & $13.274\pm0.022$ & $12.898\pm0.022$ \\
 857.39 &   30.00 & $13.316\pm0.021$ & $13.622\pm0.021$ & $13.371\pm0.022$ & $13.028\pm0.022$ \\
 892.84 &   30.00 & $13.413\pm0.021$ & $13.809\pm0.021$ & $13.473\pm0.022$ & $13.131\pm0.023$ \\
 928.38 &   30.00 & $13.528\pm0.021$ & $13.823\pm0.021$ & $13.557\pm0.022$ & $13.297\pm0.023$ \\
 963.52 &   30.00 & $13.690\pm0.021$ & $13.911\pm0.021$ & $13.681\pm0.023$ & $13.335\pm0.024$ \\
 998.76 &   30.00 & $13.744\pm0.021$ & $13.997\pm0.021$ & $13.724\pm0.023$ & $13.318\pm0.024$ \\
1034.51 &   30.00 & $13.753\pm0.021$ & $14.060\pm0.021$ & $13.836\pm0.023$ & $13.436\pm0.024$ \\
1070.46 &   30.00 & $13.886\pm0.021$ & $14.157\pm0.021$ & $13.870\pm0.023$ & $13.471\pm0.024$ \\
1106.31 &   30.00 & $13.914\pm0.022$ & $14.237\pm0.022$ & $13.932\pm0.024$ & $13.610\pm0.026$ \\
1142.56 &   30.00 & $13.935\pm0.022$ & $14.328\pm0.022$ & $13.973\pm0.024$ & $13.660\pm0.026$ \\
1177.80 &   30.00 & $14.010\pm0.022$ & $14.391\pm0.022$ & $14.064\pm0.024$ & $13.741\pm0.026$ \\
1213.45 &   30.00 & $14.156\pm0.022$ & $14.438\pm0.023$ & $14.162\pm0.026$ & $13.802\pm0.028$ \\
1249.20 &   30.00 & $14.145\pm0.022$ & $14.396\pm0.022$ & $14.179\pm0.026$ & $13.863\pm0.028$ \\
1284.84 &   30.00 & $14.305\pm0.022$ & $14.520\pm0.022$ & $14.280\pm0.026$ & $13.900\pm0.028$ \\
1320.29 &   30.00 & $14.277\pm0.022$ & $14.616\pm0.023$ & $14.419\pm0.028$ & $14.030\pm0.031$ \\
1355.93 &   30.00 & $14.332\pm0.022$ & $14.718\pm0.023$ & $14.328\pm0.026$ & $14.044\pm0.030$ \\
1391.59 &   30.00 & $14.408\pm0.023$ & $14.683\pm0.024$ & $14.362\pm0.028$ & $14.150\pm0.033$ \\
1426.92 &   30.00 & $14.455\pm0.022$ & $14.776\pm0.023$ & $14.511\pm0.028$ & $14.139\pm0.030$ \\
1462.27 &   30.00 & $14.446\pm0.023$ & $14.857\pm0.024$ & $14.535\pm0.028$ & $14.230\pm0.031$ \\
1497.71 &   30.00 & $14.728\pm0.024$ & $14.891\pm0.025$ & $14.591\pm0.032$ & $14.309\pm0.038$ \\
1532.96 &   30.00 & $14.614\pm0.022$ & $14.891\pm0.023$ & $14.629\pm0.028$ & $14.315\pm0.032$ \\
1568.30 &   30.00 & $14.637\pm0.024$ & $14.932\pm0.025$ & $14.735\pm0.032$ & $14.379\pm0.037$ \\
1603.85 &   30.00 & $14.678\pm0.024$ & $15.012\pm0.025$ & $14.796\pm0.034$ & $14.421\pm0.038$ \\
1639.29 &   30.00 & $14.734\pm0.024$ & $15.038\pm0.025$ & $14.805\pm0.032$ & $14.537\pm0.038$ \\
1674.64 &   30.00 & $14.784\pm0.025$ & $15.085\pm0.027$ & $14.886\pm0.036$ & $14.479\pm0.041$ \\
1710.48 &   30.00 & $14.804\pm0.024$ & $15.082\pm0.026$ & $14.927\pm0.034$ & $14.623\pm0.040$ \\
1745.83 &   30.00 & $14.808\pm0.024$ & $15.156\pm0.026$ & $14.895\pm0.034$ & $14.594\pm0.039$ \\
1780.96 &   30.00 & $14.932\pm0.026$ & $15.229\pm0.028$ & $15.061\pm0.039$ & $14.750\pm0.047$ \\
1816.52 &   30.00 & $15.003\pm0.024$ & $15.263\pm0.026$ & $14.966\pm0.034$ & $14.715\pm0.041$ \\
1852.06 &   30.00 & $15.028\pm0.026$ & $15.274\pm0.028$ & $15.019\pm0.037$ & $14.666\pm0.043$ \\
1887.61 &   30.00 & $14.996\pm0.028$ & $15.269\pm0.029$ & $14.994\pm0.039$ & $14.774\pm0.049$ \\
1923.25 &   30.00 & $15.004\pm0.027$ & $15.342\pm0.030$ & $15.209\pm0.043$ & $14.726\pm0.047$ \\
1958.70 &   30.00 & $15.046\pm0.028$ & $15.352\pm0.030$ & $15.073\pm0.041$ & $14.832\pm0.053$ \\
1993.94 &   30.00 & $15.079\pm0.026$ & $15.443\pm0.028$ & $15.143\pm0.039$ & $14.882\pm0.047$ \\
2029.29 &   30.00 & $15.176\pm0.028$ & $15.333\pm0.030$ & $15.197\pm0.044$ & $14.889\pm0.055$ \\
2064.93 &   30.00 & $15.157\pm0.025$ & $15.457\pm0.028$ & $15.182\pm0.036$ & $14.954\pm0.048$ \\
2100.88 &   30.00 & $15.188\pm0.028$ & $15.455\pm0.031$ & $15.371\pm0.047$ & $14.925\pm0.052$ \\
2136.02 &   30.00 & $15.222\pm0.028$ & $15.403\pm0.029$ & $15.268\pm0.042$ & $14.965\pm0.052$ \\
2171.37 &   30.00 & $15.294\pm0.028$ & $15.477\pm0.030$ & $15.338\pm0.044$ & $14.858\pm0.048$ \\
2224.64 &  150.00 & $15.229\pm0.024$ & $15.516\pm0.025$ & $15.330\pm0.031$ & $14.985\pm0.035$ \\
2401.79 &  150.00 & $15.379\pm0.025$ & $15.680\pm0.027$ & $15.479\pm0.034$ & $15.139\pm0.041$ \\
2667.83 &  300.00 & $15.549\pm0.023$ & $15.884\pm0.024$ & $15.593\pm0.028$ & $15.281\pm0.033$ \\
3023.57 &  300.00 & $15.828\pm0.024$ & $16.044\pm0.025$ & $15.799\pm0.031$ & $15.537\pm0.039$ \\
3379.09 &  300.00 & $15.934\pm0.025$ & $16.227\pm0.027$ & $15.989\pm0.034$ & $15.625\pm0.042$ \\
3734.19 &  300.00 & $15.981\pm0.026$ & $16.306\pm0.028$ & $16.071\pm0.037$ & $15.822\pm0.047$ \\
4265.78 &  600.00 & $16.260\pm0.024$ & $16.471\pm0.025$ & $16.260\pm0.032$ & $15.965\pm0.039$ \\
4976.18 &  600.00 & $16.407\pm0.026$ & $16.763\pm0.028$ & $16.496\pm0.036$ & $16.151\pm0.046$ \\
5686.59 &  600.00 & $16.660\pm0.027$ & $16.930\pm0.030$ & $16.661\pm0.040$ & $16.322\pm0.050$ \\
6396.62 &  600.00 & $16.780\pm0.028$ & $17.046\pm0.030$ & $16.853\pm0.043$ & $16.635\pm0.064$ \\
\enddata
\end{deluxetable}

\clearpage

\begin{figure}
\plotone{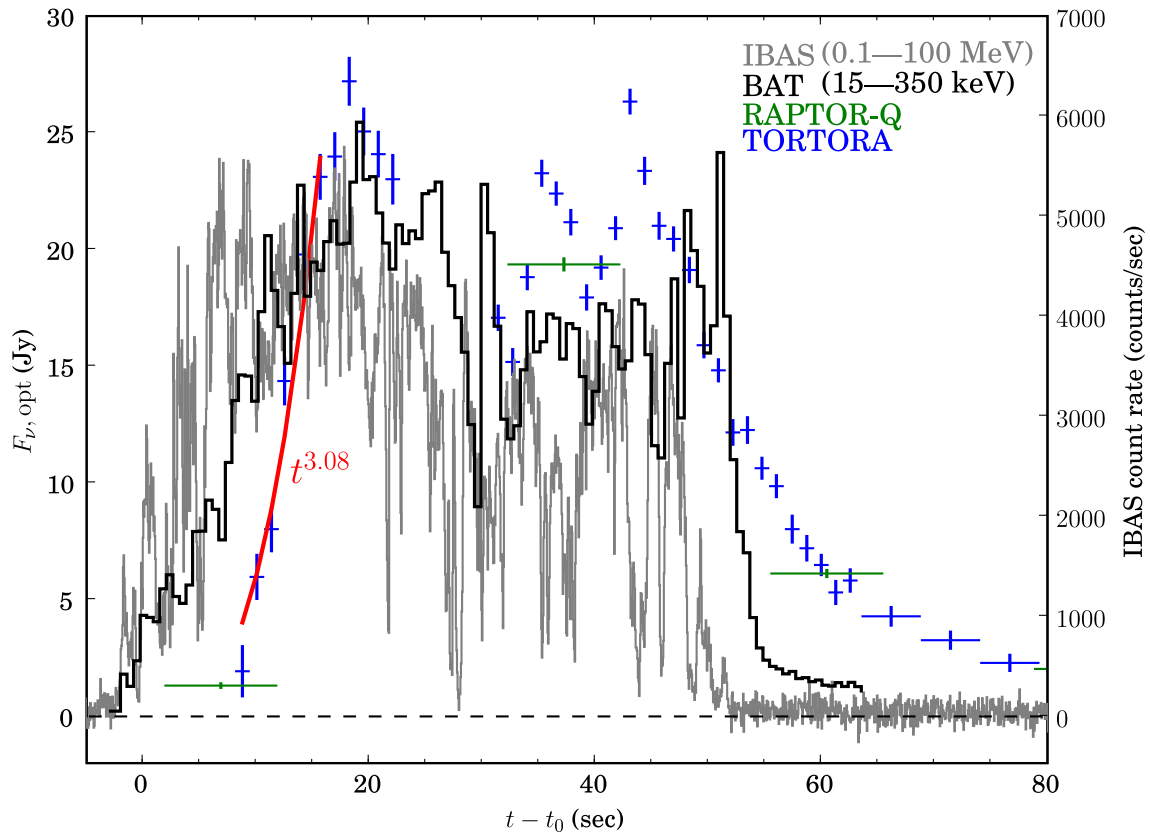}
\figcaption[]{\label{fig:gamma_lc}
A comparison of the prompt optical light curve and the hard X-ray to gamma-ray light curve of GRB 080319B.
The average gamma-ray flux shows a gradual rise to the peak value and is accompanied by extremely rapid onset
$\sim t^3$ of very strong prompt optical emission. In the synchrotron self-Compton
scenario the synchrotron spectrum peaking near the optical frequency window is initially suppressed
by self-absorption. The expanding fireball quickly becomes optically thin at low energies allowing
photons near the peak of synchrotron emissivity to escape.
}
\end{figure}

\end{document}